\documentclass[12pt,a4paper]{article}
\usepackage{epsfig} 
\usepackage{graphics}
\usepackage{subfigure}
\usepackage{a4wide}
\usepackage{amsfonts}

\title{Hopf Solitons from Instanton Holonomy}
 \author{R S Ward\footnote{email: richard.ward@durham.ac.uk}
 \bigskip
\\Department of Mathematical Sciences,  \\ University of
Durham, \\Durham DH1 3LE}

\newcommand{\ie}{{\it ie\ }}
\newcommand{\cf}{{\it cf\ }}
\newcommand{\half}{{\textstyle\frac{1}{2}}}

\newcommand{\RR}{{\mathbb{R}}}

\newcommand{\pa}{\partial}

\newcommand{\ii}{{\rm i}}

\newcommand{\s}{\sigma}

\renewcommand{\th}{\theta}

\renewcommand{\l}{\lambda}
\renewcommand{\L}{\Lambda}

\renewcommand{\O}{\Omega}

\begin{document}

\maketitle \abstract{The holonomy of an SU(2) $N$-instanton in the
$x^4$-direction gives a map from $\RR^3$ into SU(2), which provides
a good model of an $N$-Skyrmion.  Combining this map with the standard
Hopf map from ${\rm SU(2)}\cong S^3$ to $S^2$ gives a configuration
for a Hopf soliton of charge $N$.  In this way, one may define a 
collective-coordinate manifold for Hopf solitons.  This paper deals with
instanton approximations to Hopf solitons in the Skyrme-Faddeev model,
focussing in particular on the $N=1$ and $N=2$ sectors, and the two-soliton
interaction.
}

\section{Introduction}

In the study of topological solitons, an important question is whether one can
approximate the soliton interactions in terms of dynamics on a
finite-dimensional manifold $M$ of `collective coordinates'.  Not only
is this useful for understanding the classical dynamics of solitons,
but it also allows an approximate quantum theory to be constructed (by
quantizing the collective coordinates).  In cases where there are no forces
between static solitons, the moduli space of static multi-soliton
solutions is an obvious candidate for $M$; examples include the
abelian Higgs model (vortices) and the Yang-Mills-Higgs model (monopoles),
both at critical Higgs self-coupling,
in two and three spatial dimensions respectively.  There is a natural
metric on the moduli space, corresponding to the expression for the kinetic
energy of the field; and geodesics with respect to this metric give an
approximate description of the multi-soliton dynamics~\cite{M82}.

If there are inter-soliton forces, then the space of static multi-soliton
solutions has too low a dimension to serve as $M$.  In some cases (such as
the two examples mentioned above, with Higgs self-coupling close to critical),
there may be a `nearby' moduli space which will do.  But in general,
something different is needed.  One proposal (\cf\ \cite{M88}) is to take $M$
to be the union of gradient-flow (steepest-descent) curves from an
appropriate saddle-point solution.  This idea has been investigated in
several examples; one of these is the Skyrme model, where the field is a
map from $\RR^3$ to SU(2).  The gradient-flow paths cannot be found
explicitly, and obtaining them numerically is a hard (3+1)-dimensional
computational exercise.  But there appears to be a good approximation,
whereby the relevant Skyrmion configurations are obtained from SU(2)
Yang-Mills instantons in $\RR^4$ \cite{AM89, HGOA90, AM93, LMS95}.  The
same idea works for SU($N$) Skyrmions \cite{LM94, I00}; and for some
lower-dimensional field theories (see, for example, \cite{S92, S95}).
The instantons are known explicitly, and the Skyrme field is set equal to
the holonomy of the instanton connection in the imaginary-time direction.
In general, the holonomy has to be computed numerically; but this involves
solving ordinary (rather than partial) differential equations, so is more
straightforward.

There is no obvious reason why the holonomy of instantons (in one system)
should provide a good
approximation to solitons (in a completely different system).  But there are
various features which make the construction a natural one.  First, an
$N$-instanton produces an $N$-Skyrmion configuration; in other words,
the topological classification is preserved.  Secondly, most symmetries
of the instanton feed through into symmetries of the Skyrmion (some symmetry
may be lost because of the choice of imaginary-time direction along which to
compute the holonomy).  The first, and to some extent the second, of these
features are also present in this paper, which investigates how instanton
holonomy can provide Hopf-soliton configurations.  In particular, we shall
see that instantons give a reasonably good approximation in the $N=1$ and
$N=2$ sectors.

Hopf solitons are topological solitons in systems involving a field
$\phi : \RR^3\to S^2$ (or $\phi : \RR^{3+1}\to S^2$, if one includes
time-dependence).   Such a field configuration is classified topologically
by its Hopf number $N \in \pi_3(S^2)$.  There are various choices for the
dynamics of the solitons, depending on which application one has in mind.
For example, if $\phi$ represents the local magnetization in a ferromagnet,
then the appropriate equation of motion is the Landau-Lifshitz equation;
for a study of the corresponding evolution of Hopf solitons, see \cite{C99}.
The present paper deals with the Skyrme-Faddeev system \cite{F75, VK79, KR82},
where the dynamics is determined by an expression of the form
$E = \int\bigl[(d\phi)^2+F^2\bigr]\,d^3x$ for the energy of $\phi$.
The second term in this expression is a Skyrme term, which serves to
stablilize the size of the soliton.  In the last few years, the Hopf solitons
arising in this system have been the subject of considerable study, mostly
involving numerical simulation
\cite{FN97, GH97, BS98, BS99, HS99, W99, W00, HS00}.

Let us represent $\phi$ as a unit 3-vector field
$\vec{\phi} = (\phi^1,\phi^2,\phi^3)$ depending on the spatial coordinates
$x^j = (x,y,z)$.  The boundary condition is
$\vec{\phi}\to\vec{n}:=(0,0,1)$ as
$r\to\infty$, where $r^2 = x^2+y^2+z^2$; this allows us to think of $\phi$
as being defined on compactified space $S^3$, and hence its Hopf number
$N$ is well-defined.  We may visualize $N$ as a linking number: if $p$ and $q$
are two generic points on the target space $S^2$, then the inverse images
$\phi^{-1}(p)$ and $\phi^{-1}(q)$ are curves in $\RR^3$,
and the curves are linked $N$ times around each other.  One such curve,
namely the inverse image of $(0,0,1)$, includes the point at infinity in
$\RR^3$.  In general, we shall visualize a soliton configuration in terms
of the inverse image $\phi^{-1}(s)$ of the point $s=(0,0,-1)$ (the
antipode of the boundary value $\vec{n}$).

The energy $E$ of the static field $\vec{\phi}(x^j)$ is taken to be
\begin{equation} \label{energy}
 E = \frac{1}{32\pi^2}
       \int \bigl[ (\partial_j\vec{\phi})\cdot(\partial_j\vec{\phi})
         + F_{jk} F_{jk} \bigr] \, d^3x,
\end{equation}
where $F_{jk} = \vec{\phi}\cdot
             (\partial_j\vec{\phi})\times(\partial_k\vec{\phi})/2$.
There is a lower bound on the energy which is proportional to $N^{3/4}$;
and if space is allowed to be a three-sphere,
then there is an $N=1$ solution with $E=1$ \cite{W99}; this is the reason
for the factor of $1/32\pi^2$ in (\ref{energy}).  So one expects, with this
normalization, that $E$ satisfies the bound $E\geq N^{3/4}$; but only the
weaker bound $E\geq cN^{3/4}$ with $c = 2^{-3/2} 3^{3/8}$ has been proved
\cite{VK79, KR82}.

No configuration with $N\neq0$ can be spherically-symmetric \cite{KR82},
but axial symmetry is allowable.  The minimum-energy solutions for $N=1$ and
$N=2$ are indeed axially-symmetric \cite{GH97}.  We say that a configuration
$\phi^a$ is symmetric about the $z$-axis if
\[
  \phi^1 + \ii\phi^2 = (x+\ii y)^m\, u,
\]
where $u$ and $\phi^3$ are functions of $z$ and $r$, and where $m$ is an
integer.  Then (\cf\ \cite{GH97}), $m$ divides $N$; so for $N=1$
we must have $m=1$, but for $N=2$ we can have either $m=1$ or $m=2$.
As far as instanton holonomy is concerned, axially-symmetric Hopf-soliton
configurations are obtained from rotationally-symmetric instantons.
The latter objects are of particular interest because they correspond to
hyperbolic monopoles \cite{A84, N86}; in that context, $m$ is the
asymptotic norm of the Higgs field (note that $m/2$ is denoted $p$ in
\cite{A84} and $n$ in \cite{N86}).

The  minimum-energy $N=1$ Hopf soliton has energy $E=1.224$ \cite{W00}.
If one chooses its axis of symmetry to be the $z$-axis as above, then
$\phi^{-1}(s)$ is a circle (of radius about 0.8) centred on the $z$-axis.
There are six obvious degrees of freedom which one
may use as collective coordinates: the location of the centre of the
circle in space (three), the direction of the axis of symmetry (two),
and a U(1) phase.  The {\em standard orientation} has
$\phi^1 + \ii\phi^2 = (x+\ii y) u$ with $u$ real-valued; a phase rotation
by $\chi$ takes this to $\phi^1 + \ii\phi^2 = (x+\ii y) u \exp(\ii\chi)$.
From a distance, the soliton resembles a pair of scalar dipoles, orthogonal
to each other and to the axis of symmetry \cite{W00}.  So its orientation
corresponds to a choice of frame in 3-space; if we fix the centre of the
soliton, then the manifold of the three remaining collective coordinates
(direction of axis plus phase) is SO(3).

In the $N=2$ case, therefore, it is natural to look for a
collective-coordinate space which twelve-dimensional. The force between two
solitons depends on their relative orientation, and can be understood
(for widely-separated solitons) in terms of the interaction between the
dipoles referred to above \cite{W00}.  The minimum-energy configuration
in the $N=2$ sector is axially-symmetric with $m=2$, and has energy
$E=2.00$ \cite{GH97}.  But there is also a local minimum, axially-symmetric
with $m=1$, and with energy $E=2.26$ \cite{W00}.  The latter is the minimum
in an `attractive channel' in which the two solitons are co-axial and in phase;
for example, two solitons centred on the $z$-axis at $z=\pm c$, each
in the standard orientation (this is referred to as channel A in \cite{W00}).
All the configurations in this channel are axially-symmetric with $m=1$.
By contrast, the most attractive channel (\ie\ the relative orientation
of the two solitons for which the force between them is maximally attractive)
has the symmetry axis of each soliton being orthogonal to the line joining
them (this is discussed in more detail in \cite{W00}; note that
figures 2 and 3 in that paper should be swapped, their captions remaining
unchanged).

The next section deals with instanton holonomy and the approximate $N=1$
soliton.  We then study the $N=2$ case, investigating various two-soliton
configurations, and the extent to which instanton holonomy can give a
suitable twelve-dimensional space of collective coordinates.


\section{Instanton holonomy and the one-soliton}

One constructs approximate Hopf-soliton configurations, with Hopf number
$N$, as follows.  The procedure is simply to apply the standard Hopf map
to approximate Skyrmion configurations.  Let
$X\leftrightarrow X^\mu = (X^0,X^1,X^2,X^3) = (t,x^j)$ denote the standard
coordinates on Euclidean 4-space $\RR^4$, and let $A_\mu$ be an SU(2)
gauge potential on $\RR^4$ with topological charge $N$.  For a fixed
$x^j\in\RR^3$, let $U(x^j)$ denote the holonomy in the $t$-direction,
namely
\[
 U(x^j) = {\mathcal T} \exp
      \biggl[-\int_{-\infty}^{\infty}A_t(t,x^j)\,dt\biggr],
\]
where ${\mathcal T}$ denotes time-ordering.  In practice, one obtains $U(x^j)$
by computing the $2\times2$ matrix solution $M(t,x^j)$ of the system
\begin{equation}\label{ODE}
 \frac{\pa M}{\pa t} = - A_t M
\end{equation}
with the initial condition $M(-\infty,x^j) = 1$, and then setting
$U(x^j) = M(\infty,x^j)$.  This function $U(x^j)$ takes values in the
gauge group ${\rm SU(2)} \cong S^3$.  It is therefore a Skyrmion configuration
\cite{AM89}.  Now applying the Hopf map from $S^3$ to $S^2$ gives an
$S^2$-valued
field $\vec{\phi}(x^j)$.  Explicitly, we get from the $2\times2$ matrix
$U_{ab}(x^j)$ to the stereographic projection
\[ 
 W = \frac{\phi^1+\ii\phi^2}{1+\phi^3}
\]
by setting $W(x^j) = U_{21}(x^j)/U_{11}(x^j)$.  If the gauge potential
$A_\mu$ decays suitably as $|X|\to\infty$ (which will be the case in what
follows), then $W(x^j)$ satisfies the required boundary condition
$W(x^j)\to0$ as $r\to\infty$, and it has Hopf number $N$.

In the Skyrme system, the energy is invariant under isospin transformations
\begin{equation}\label{isospin}
 U \mapsto \L^{-1} U \L,
\end{equation}
where $\L\in{\mathrm SU(2)}$ is constant.  So $\L$ provides three additional
parameters, which in the Skyrme case do not affect the energy.  Since $\L$
and $-\L$ determine the same transformation, this parameter space is an
SO(3).  The Hopf map, however, breaks the symmetry; so the transformations
(\ref{isospin}) have some significance (in general).  We shall see later that
these additional parameters are necessary in the $N=2$ case.  This amounts
to using a family of Hopf maps from SU(2) to $S^2$, rather than just one.

The construction above works for any gauge field; but the idea is that
instantons lead to particularly relevant configurations.  There is a
simple formula (\cf\ \cite{JNR77}) for the instanton solutions with $N=1,2$;
in particular, $A_t$ has the form
\[
  A_t = \half\ii\,\O^{-1}(\pa_j\O) \s_j,
\]
where $\s_j$ denotes the Pauli matrices, and where
\begin{equation} \label{Omega}
 \O(X) = \sum_{a=1}^{N+1} \frac{\l_a^2}{|X-X_a|^2}.
\end{equation}
Here the $X_a$ are distinct points in $\RR^4$.  Although $A_t$ has poles at
these points, the poles are removable; and the resulting Hopf configuration
$\vec{\phi}(x^j)$ is smooth on $\RR^3$.  This ansatz produces an $N$-instanton
solution for any $N\geq1$, and for $N=1$ and $N=2$ it produces all the
instantons in the corresponding topological sectors.

Let us consider, first, the $N=1$ sector.  A special case of (\ref{Omega})
is the 'tHooft expression; for this, one takes the limit
$\l_2=|X_2|\to\infty$, giving
\begin{equation} \label{tHooft}
 \O(X) = 1 + \frac{\l_1^2}{|X-X_1|^2}.
\end{equation}
The formula (\ref{tHooft}) depends on the five real parameters
$(\l_1,X_1^{\mu}) = (\l_1,t_1,x_1,y_1,z_1)$.
Without loss of generality, we may set $t_1=0$ (since we are integrating
over $t$).  If we centre the soliton in 3-space by choosing $x_1=y_1=z_1=0$,
then only one parameter remains: the scale factor $\l=\l_1$.

In this case, we can compute the holonomy analytically \cite{AM89}, and
we then obtain the Hopf-soliton configuration
\begin{equation} \label{1-sol}
 W = \frac{x+\ii y}{z-\ii r\cot f(r)},
\end{equation} 
with the profile function $f(r)$ being
\begin{equation} \label{profile}
 f(r) = \pi\left[ 1 - r/\sqrt{r^2+\l^2} \right].
\end{equation} 
So (\ref{1-sol}), (\ref{profile}) gives a one-parameter family of Hopf-soliton
configurations.  This is the analogue of the `hedgehog'
configuration in the Skyrme model, and in fact the expression for its energy
is exactly the same functional of $f$ in the two systems
(assuming an appropriate choice of coupling constants).  So we already
know from the Skyrme case (\cf\ \cite{AM89, AM93}) that the energy
of the configuration is $E(\l) = 0.428\,\l + 0.903/\l$; and this
has a minimum value of $E=1.243$ when $\l=1.45$.

In the Skyrme model, the actual 1-Skyrmion is a hedgehog (spherically
symmetric), with profile $f(r)$
a slightly deformed version of (\ref{profile}), and energy $E=1.231$.
But the analogous statement is not true for Hopf solitons; in other words,
the minimum-energy Hopf soliton does not quite have the form (\ref{1-sol}).
The actual solution (minimum-energy configuration in the $N=1$ sector)
is a slight deformation of a hedgehog; it has energy $E=1.224$,
as mentioned previously.  But the instanton-derived configuration
is nevertheless sufficiently close to the actual solution to be a useful
approximation (in particular, its energy is only $1.5\%$ higher than
the actual minimum).   Note that (\ref{1-sol}) is in the standard orientation;
and that the locus of points where $\phi^3=-1$ is a ring in the
$xy$-plane, with centre $x=y=z=0$ and radius $r\approx0.84$ ($\phi^3=1$ is
the $z$-axis).  For large $r$, we have
\[
  \phi^1+\ii\phi^2 \approx \alpha\frac{x+\ii y}{r^3},
\]
where $\alpha$ is a constant; so $\phi^1$ and $\phi^2$ resemble a pair of
orthogonal dipoles.

The more general expression (\ref{Omega}) contains ten parameters
$(\l_1,\l_2,X_1^{\mu},X_2^{\mu})$.  Of these, three determine the location
\begin{equation} \label{location}
 p^j = \frac{\l_2^2 x_1^j - \l_1^2 x_2^j}{\l_1^2 + \l_2^2}
\end{equation} 
of the soliton in 3-space; one determines the scale
\begin{equation} \label{scale}
 \l = \frac{\l_1 \l_2}{\l_1^2 + \l_2^2}|X_2 - X_1|,
\end{equation} 
which is set to $1.45$ for minimum energy; and three, forming an SO(3),
determine the direction of the line $L$ in $\RR^4$ from $X_1$ to $X_2$, which
in turn determines the phase and the direction of the symmetry axis of the
soliton.  The remaining three parameters have no effect, and can be
eliminated by (say) setting  $\l_1 = \l_2 = 1$ and $t_2 + t_1 = 0$.  
So we get, as required, a six-parameter family of soliton configurations.
For example, the choice $x_1^j = x_2^j = 0$, $\l_1 = \l_2 = 1$ and
$t_2 = -t_1 = 1.45$ gives a soliton at the origin, with the standard
orientation.  In this case, the line $L$ from $X_1$ to $X_2$ is
in the $t$-direction.  If we rotate $L$ by taking $t_2 = -t_1 = 1.45\cos\th$
and $z_2 = -z_1 = 1.45\sin\th$, then we get the standard configuration
rotated by an angle $2\th$ about the $z$-axis (a phase rotation).
And rotating $L$ in (say) the $x$-direction by an angle $\th$,
has the effect of rotating the spatial axis in that direction by an angle
$2\th$; for example, taking $t_2 = -t_1 = x_2 = -x_1 = 1.45\cos(\pi/4)$
gives a soliton with symmetry-axis the $x$-axis rather than the $z$-axis.

Two remarks end this section.  First, there is another (equivalent)
way of introducing the orientation degrees of freedom in this $N=1$ case,
namely by using the SO(3) transformations (\ref{isospin})
applied to the standard-orientation soliton (\ref{1-sol}).  Secondly,
it follows from (\ref{scale}) that $\l \leq |X_2 - X_1|/2$;
so (given that we want $\l=1.45$), the distance between the two poles
$X_2$ and $X_1$ in $\RR^4$ has to be at least $2\times1.45$.


\section{Axisymmetric $m=1$ two-soliton configurations}

Let us now take $N=2$ in (\ref{Omega}), so that we have fifteen parameters
$(\l_a,X_a^{\mu})$.  Of these parameters, one (the overall $\l$-scale) is
irrelevant, since it does not affect $A_{\mu}$.  For instantons, there
is an additional degeneracy \cite{JNR77} which will be referred to here
as the {\em porism} freedom \cite{AM93}: the three poles lie
on a circle (or line) in $\RR^4$, and if the poles move around this circle in
a certain way, than the only effect on the instanton is to induce a gauge
transformation.  So the space of 2-instantons is 13-dimensional.  The
effect of the porism freedom on a holonomy-generated Skyrmion is either
trivial, or it induces an isospin rotation (\ref{isospin}); either way, the
energy does not change.  But the effect on the Hopf configuration is, in
general, non-trivial (and can, for example, alter the relative orientation
and hence the force between the two solitons).  This effect is contained
in the extra degrees of freedom (\ref{isospin}); so let us remove the porism
freedom, while retaining, for the time being, these three extra parameters.
Note that one of them corresponds to a global phase rotation
$W\mapsto W\exp(\ii\chi)$, and does not change the soliton energy.
We can also remove a
parameter by translating in $t$ (since we are integrating in that direction).
So we are left with a 15-dimensional space of Hopf-soliton configurations.

This gives enough freedom to generate the twelve-dimensional space of two
solitons with all possible (well-separated) positions and orientations.
To see this, we may argue as follows (\cf\ \cite{AM93}).  The instanton,
and hence the soliton, are determined by (\ref{Omega}) with $N=2$.  First, note
that if $\l_3\gg\l_1$ and $\l_3\gg\l_2$, and if $x^j_1$ and $x^j_2$ are
well-separated, then in a neighbourhood of $X_1$ the $\l_2$-term is
negligible; in view of (\ref{location}), we then have a soliton located at
$x^j_1$ with orientation determined by the direction of the line from
$X_3$ to $X_1$.  Call this `S1'.  Similarly, there is a soliton (`S2')
at $x^j_2$.  Now suppose in addition that $\l_2\gg\l_1$, with $X_2$
being much closer to $X_3$ than $X_1$.  For example, take $|X_3-X_2|=c$ and
$|X_3-X_1|=c^2$ where $c$ is large.  Choose $x^j_1$ and $x^j_2$ to be at
the desired locations of the two solitons (with, say, $t_1=t_2=0$).
The orientation of S1 is determined by the direction of the line from
$X_3$ to $X_1$; but this is essentially fixed (since $X_2$ is already fixed,
and $X_3$ is relatively close to $X_2$).  However, we also have the freedom
(\ref{isospin}), so we can use this to bring S1 to its desired orientation.
Finally, adjust $X_3$ by moving it around the 3-sphere of radius $c$ in
$\RR^4$, so that S2 has the desired orientation (this being
determined by the direction of the line from $X_3$ to $X_2$).
So we have two solitons with pre-determined positions and orientations.
The $\l_a$ are chosen so that each soliton has the correct scale;
in fact, if we take $\l_1=1$, then from (\ref{scale}) we get
$\l_2\approx c$ and $\l_3\approx c^2/1.45$.  The twelve parameters
$x^j_1$, $x^j_2$, $\L$ and $X_3$ (with $|X_3-X_2|^2=c^2=|x^j_1-x^j_2|$)
are collective coordinates for the two well-separated solitons.

For the remainder of this section, let us look at the
special case where one has rotational symmetry with $m=1$.
This corresponds to the poles $X_a^{\mu}$ all lying in the $tz$-plane
in $\RR^4$, and so $x_a=y_a=0$.  Thus there are nine parameters $t_a$,  $z_a$
and $\l_a$; removing the degeneracy by setting $t_3=0$ and $\l_1=1$ leaves
seven significant parameters. In addition, we may set $z_3=0$ by translating
in $z$, so that leaves a six-parameter family of soliton configurations.
The assumption of axial symmetry means that the additional parameters
(\ref{isospin}) are not relevant here; but we do still have the porism freedom.

The simplest choice for these remaining parameters is to have the poles
$X_a^{\mu}$ all lying on the $t$-axis in $\RR^4$, \ie\ $z_a=0$.
This corresponds to an $N=2$ hedgehog Skyrmion configuration \cite{AM93};
and so we can, as before, use the numerical results from the Skyrme case.
The minimum hedgehog energy $E=3.711072$ occurs for
\begin{equation} \label{Omega2}
 \O(X) = \frac{\l^2}{r^2+t^2} + \frac{1}{r^2+(t-c)^2} + \frac{1}{r^2+(t+c)^2},
\end{equation}
with the parameter-values $\l=5.6580$ and $c=9.4808$.  (The results in
\cite{AM93} are given for the 'tHooft form of $\O$, but section 6 of that
paper gives the formulae which enables one to convert to the form
(\ref{Omega2}).  See also \cite{Wal95}.)   In fact, if we set
$\l=\sqrt{0.3784\,c^2-2}$, then the configuration and its energy have a rather
weak dependence on $c$, as long as $c$ is large enough: for $3.5<c<\infty$,
the energy $E$ is within $0.2\%$ of its minimum.  The corresponding
Hopf-soliton configuration has the form (\ref{1-sol}), and may be
visualized as a pair
of concentric rings in the $z=0$ plane; these are located where
$f(r) = \pi/2$ and $f(r) = 3\pi/2$, so in this case the rings have radii
$r\approx0.4$ and $r\approx1.8$.  As before, this configuration is not a
solution of the field equations, but it is close to one; a numerical simulation
which starts at this configuration and moves down the energy gradient,
comes to rest at a very similar configuration with energy $E=3.50$. 
It is therefore reasonable to conjecture that there is an unstable
stationary point with this energy.  This is the analogue of the unstable
spherically-symmetric two-Skyrmion.

Within our six-parameter family (of axially-symmetric configurations),
the porism freedom is a zero-mode of this configuration (it is energy-neutral);
there are three
energy-increasing modes and two energy-decreasing (negative) modes.
The latter were described, in the Skyrmion context, in \cite{AM93}; they
are as follows.  Throughout, we keep $X_3$ fixed at the origin, \ie\
keep $t_3 = z_3 = 0$.

In the hedgehog configuration, the three poles are arranged collinearly, with
$z_a=0$, $t_3=0$ and $t_2 = -t_1=c$.
The first negative mode corresponds to rotating $X_2$ clockwise and
$X_1$ anticlockwise about the origin in the $tz$-plane. The weights
remain unchanged.  In other words, we take
\[
 t_2=-t_1=S\cos\th, \quad z_2=z_1=S\sin\th,
   \quad \l_1 = \l_2 = 1, \quad \l_3 = \sqrt{0.3784\,S^2-2},
\]
for some fixed value of $S$,
with $|\th|<\pi/2$ ($\th$ cannot reach $\pi/2$, for then
$X_1$ and $X_2$ would coincide).  The corresponding holonomy-generated soliton
configurations have been computed numerically, for $S=4$.  One expects this
mode to correspond to moving both rings in the positive $z$-direction,
as well as changing their relative phase;  and this is indeed what one sees,
with the energy of the configuration decreasing to $E=2.9$
(for $\th=0.6\,\pi/2$) and then increasing as $\th$ increases further,
and as the poles $X_1$ and $X_2$ approach each other.
Roughly speaking, this mode corresponds (for small $\th$) to changing
the relative phase of the two rings, without changing their position
(except for an overall translation in $z$).

The other negative mode is more interesting.  In terms of the poles in the
instanton ansatz, it is a rigid rotation about the origin in the $tz$ plane,
by an angle $\th$; so the arrangement of poles remains collinear.
In other words, we take
\[
 t_2=-t_1=S\cos\th, \quad z_2=-z_1=S\sin\th,
   \quad \l_1 = \l_2 = 1, \quad \l_3 = \sqrt{0.3784\,S^2-2}.
\]
In this case, $\th$ is unrestricted; we expect that for large $S$ and for
$\th = \pi/2$, we will have two separated solitons, in phase, at
$z\approx\pm S$, and this is indeed what happens.
The results of a numerical computation for $S=4$  are summarized in Figure 1.
\begin{figure}[bt]
\begin{center}
\subfigure[Energy as function of $\th$]{
\includegraphics[scale=0.4]{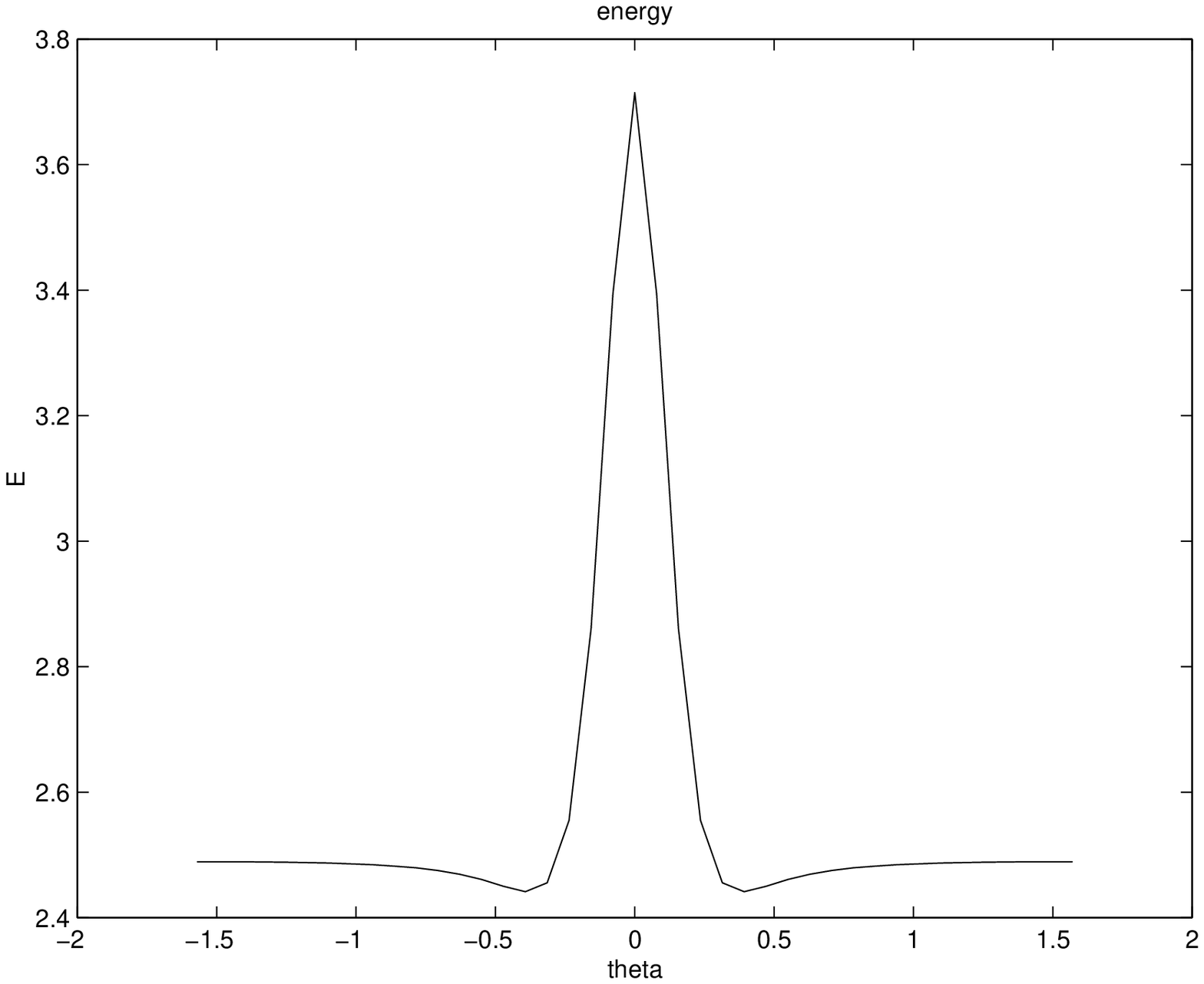}
}
\quad
\subfigure[$\th=0$: rings in $xy$-plane]{
\includegraphics[scale=0.4]{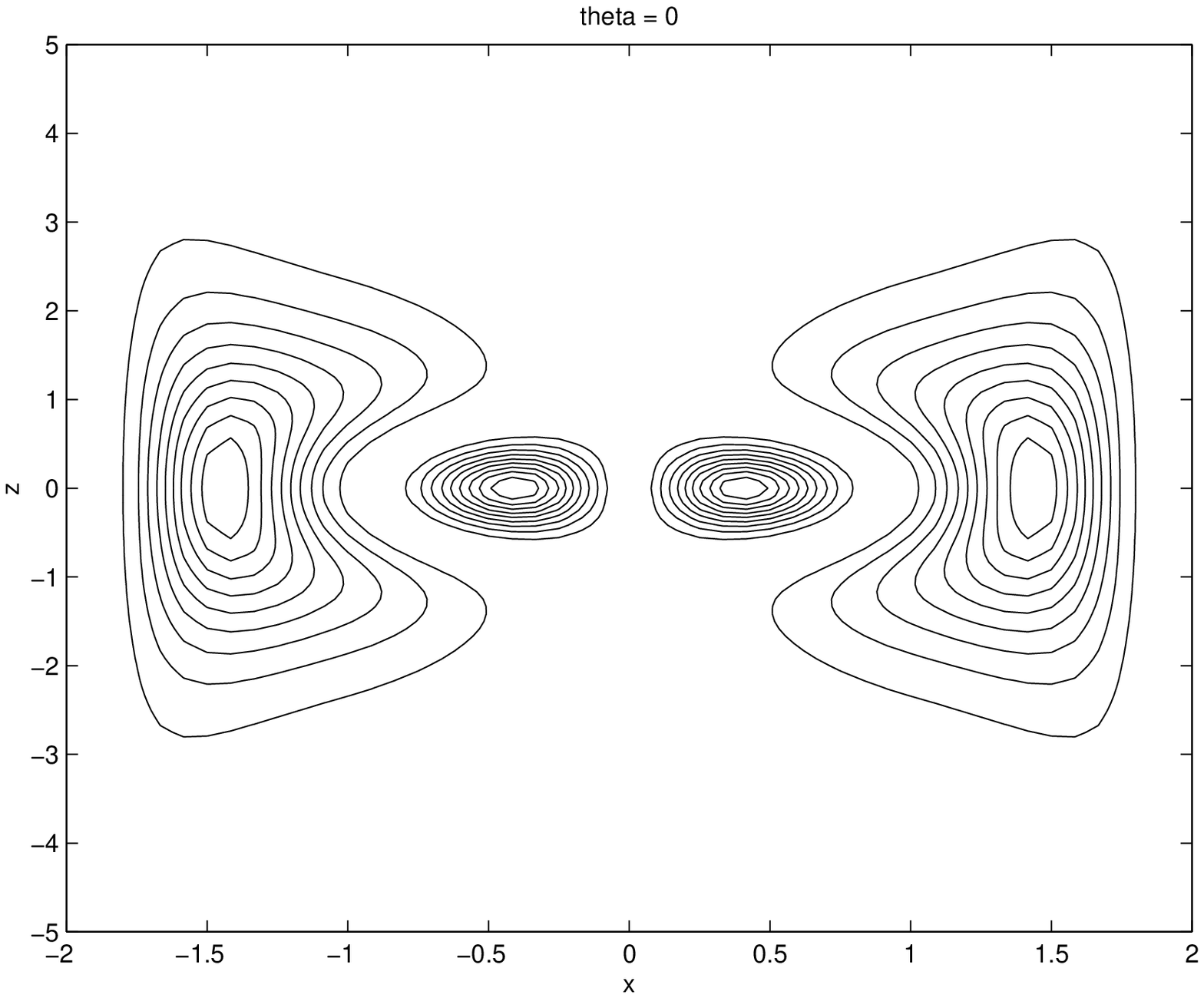}
}
\quad
\subfigure[$\th=0.16\,\pi/2$: merged rings]{
\includegraphics[scale=0.4]{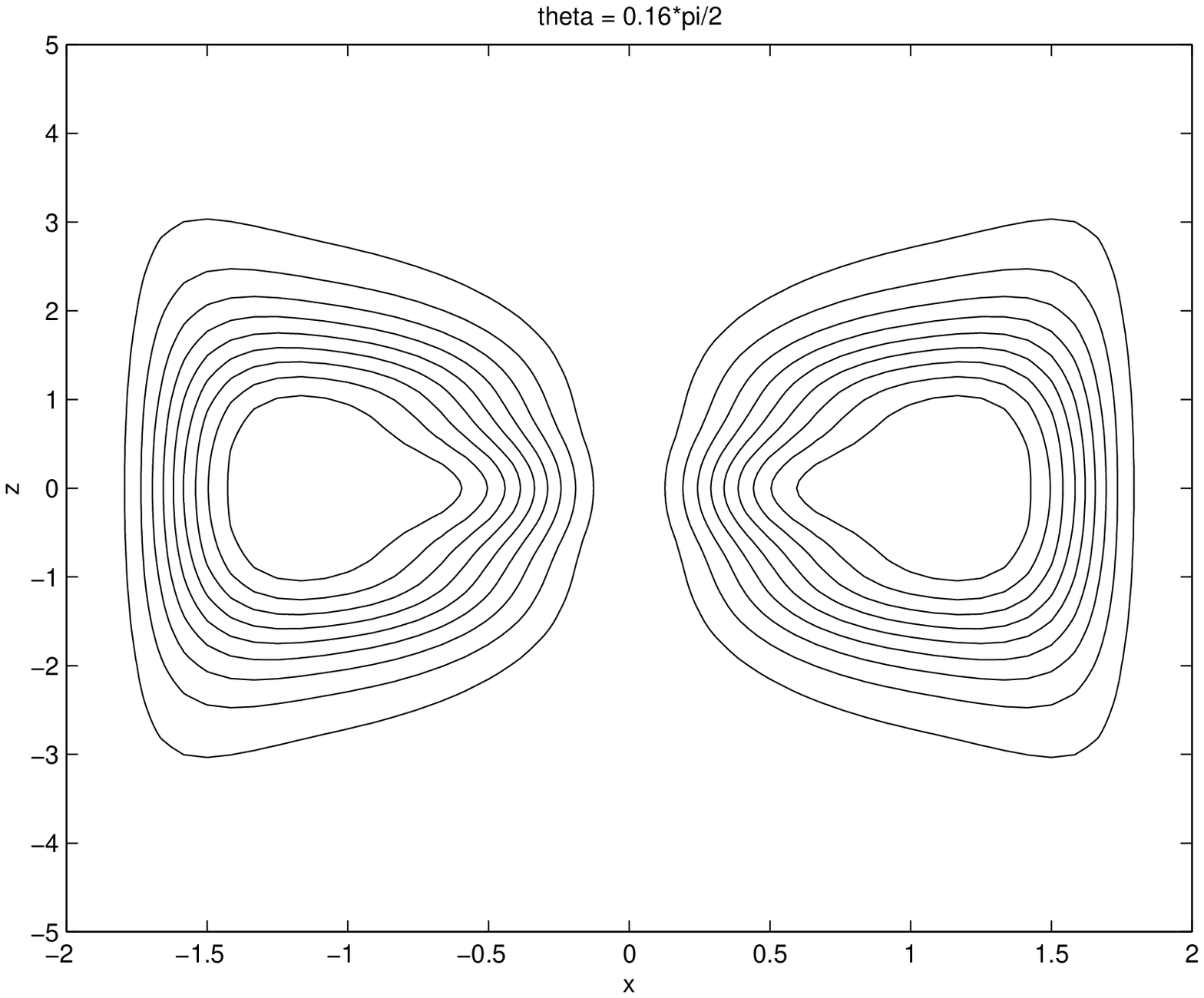}
}
\quad
\subfigure[$\th=\pi/2$: separated solitons]{
\includegraphics[scale=0.4]{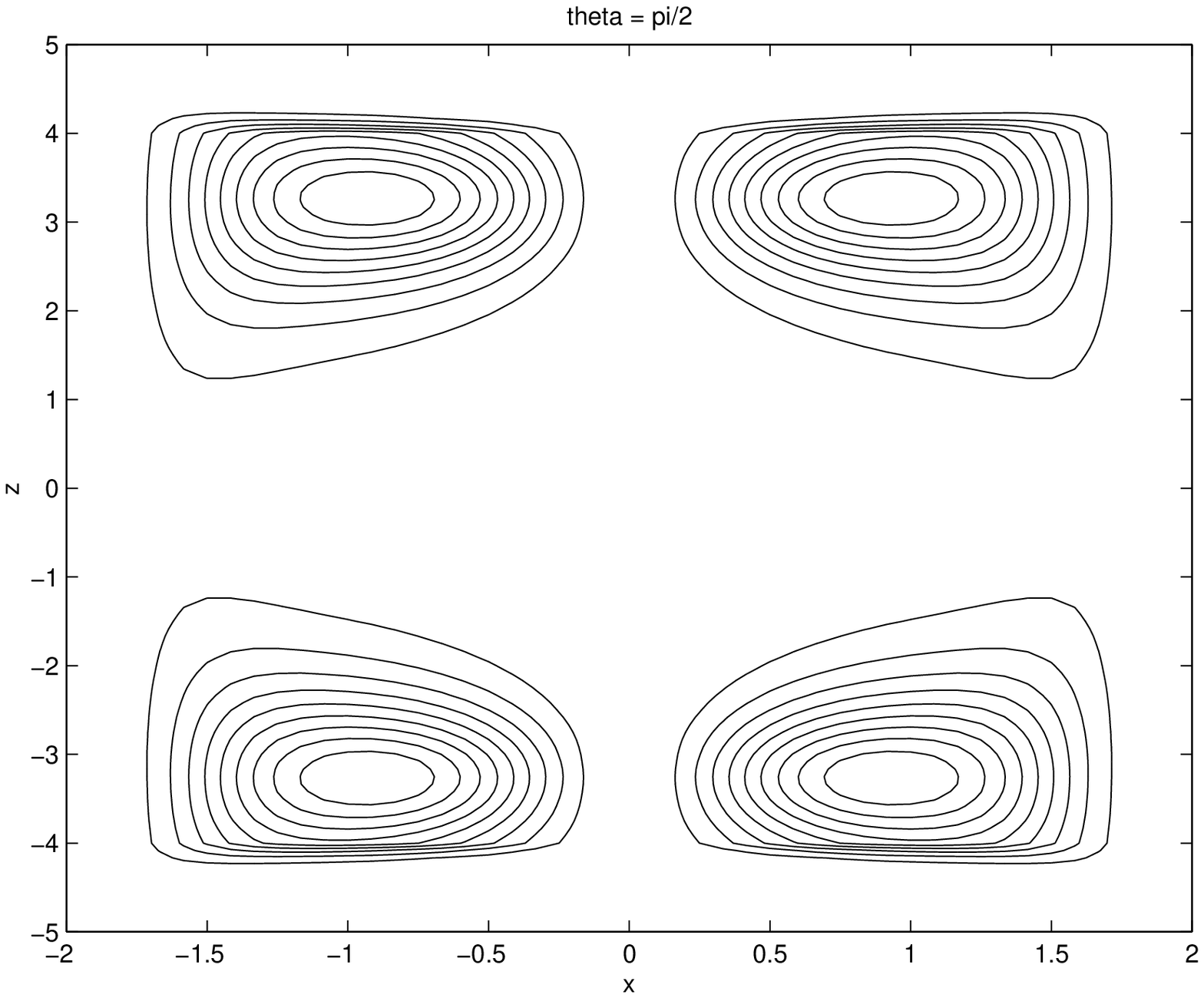}
}
\caption{A one-parameter family of soliton configurations, interpolating
   between a double ring in the $xy$-plane (with energy $E=3.71$) and
   two separated concentric solitons (with energy $E=2E_1=2.48$).
   The energy is plotted in (a), as a function of $\th$.  The other three
   subfigures are contour plots of $\phi^3$, in the $xz$-plane.\label{fig1}}
\end{center}
\end{figure}
When $\th=0$, we have two concentric rings in the $xy$-plane, as described
previously (Figure 1b).  As $\th$ increases, the rings move towards each
other (remaining in the $xy$-plane), and the energy decreases (steeply).
The rings coalesce (Figure 1c), and then begin to separate along the $z$-axis:
we now have two rings of equal radius, each in the standard configuration,
at $z=\pm p$ (Figure 1d).
The energy reaches a minimum (in this family) of $E=2.44$ when
$\th=0.25\,\pi/2$.  As $\th$ increases further towards $\pi/2$,
the separation $2p$ increases, and the energy approaches the asymptotic
value $E=2E_1=2\times1.24$ of two widely-separated 1-solitons.

Let us now keep the collinear arrangement of poles, \ie\
\begin{equation} \label{chana1}
 t_2=-t_1=S\cos\th, \quad z_2=-z_1=S\sin\th,
   \quad \l_1 = \l_2 = 1,
\end{equation}
and minimize over $S$, $\th$ and $\l_3$.  Within this three-parameter family,
there is a local minimum $E=2.42$ of the energy when $S=2.6$, $\l_3 =0.71$
and $\th=\pi/2$ (so the poles all lie on the $z$-axis).
This is not quite a minimum within our six-parameter family of axisymmetric
solitons --- the energy of the configuration generated by (\ref{chana1}) can be
reduced by increasing $t_3$ away from zero, which corresponds
to changing the relative phase of the two co-axial solitons.  However,
the dependence of $E$ on $t_3$ is not very strong, and it is reduced by less
than $1\%$; the minimum $E=2.41$ is reached when $t_3\approx0.6$.
The actual minimum
energy in this axially-symmetric class is $E=2.26$, as noted earlier.
The instanton-generated soliton configuration described above looks very
similar to this static solution, but its energy is about $7\%$ too high.


\section{Other two-soliton configurations}

This section deals with further aspects of the two-soliton parameter space.
let us begin by considering the lowest-energy solution, which is axisymmetric
with $m=2$; its energy is $E=2.00$ \cite{GH97, W00}.  To approximate
it, we use the holonomy of an appropriate rotationally-invariant instanton;
this also models the minimal-energy two-Skyrmion \cite{HGOA90, AM93}.
The instanton poles are chosen to have equal weights $\l_a=1$, and to lie at
the vertices of an equilateral triangle in the $xy$ plane (the axis of
symmetry will then be the $z$-axis).  In other words, we may take
$t_a = z_a = 0$, and
\begin{equation} \label{equilat}
  x_3 = 0, \quad y_3 = -S, \quad x_2 = -x_1 = \sqrt{3}\,S/2,
     \quad y_2 = y_1 = S/2,
\end{equation}
where $S$ is a positive constant.  The Hopf configurations corresponding
to this one-parameter family of instantons resemble a ring in the
$xy$-plane, with radius determined by $S$; this is where
$\phi(s)=(0,0,-1)$.  Because $m=2$, the locus $\phi^{-1}(0,0,1)$ actually
consists of two copies of the $z$-axis, so the linking number is indeed $N=2$.
When $S=1.72$, the energy of the family (\ref{equilat}) attains a
minimum value $E=2.08$, which is $4\%$ above the true minimum.
\begin{figure}[tb]
\begin{center}
\subfigure[Energy as function of $\l_3$]{
\includegraphics[scale=0.4]{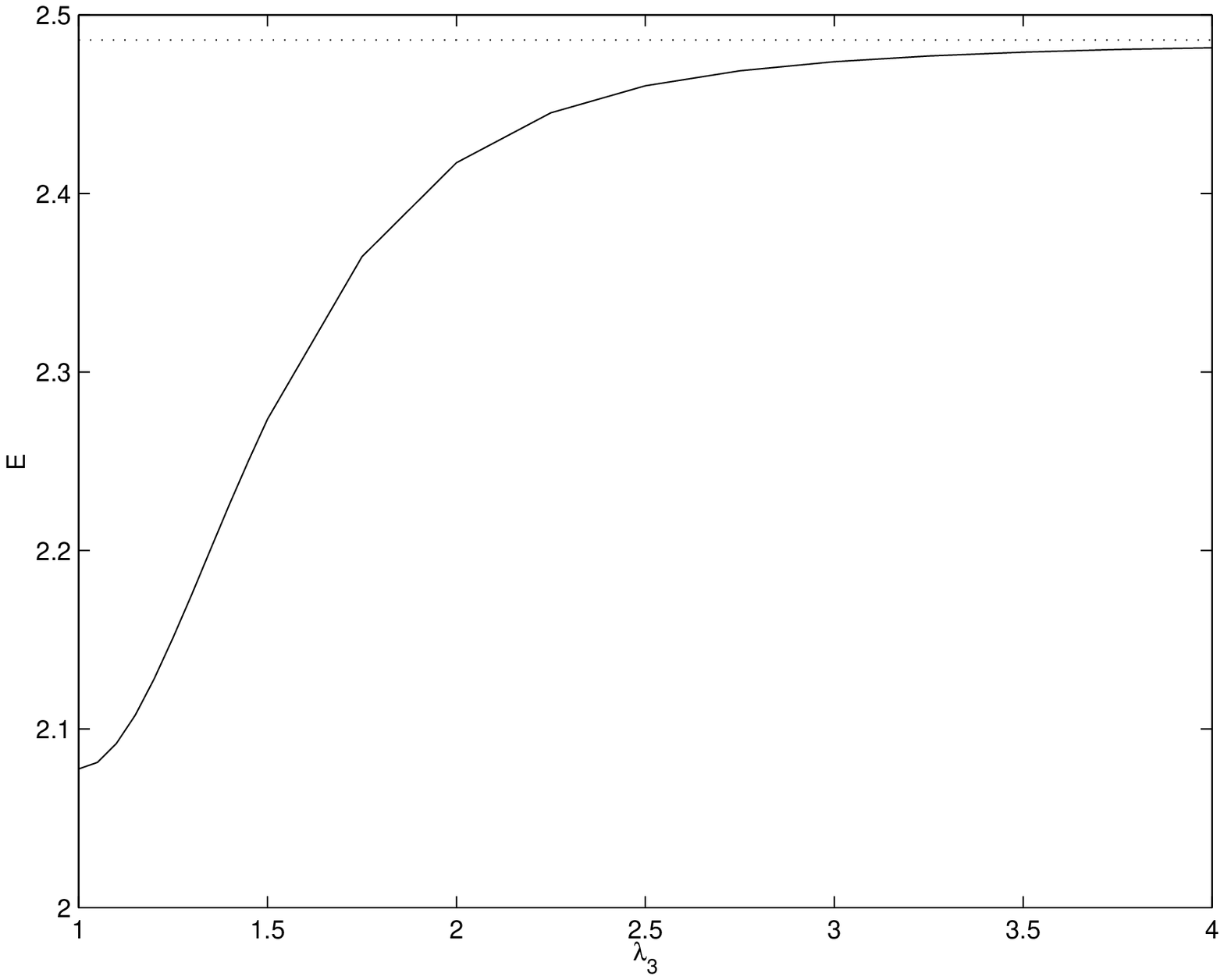}
}
\quad
\subfigure[$\l_3=4$: two separate solitons]{
\includegraphics[scale=0.4]{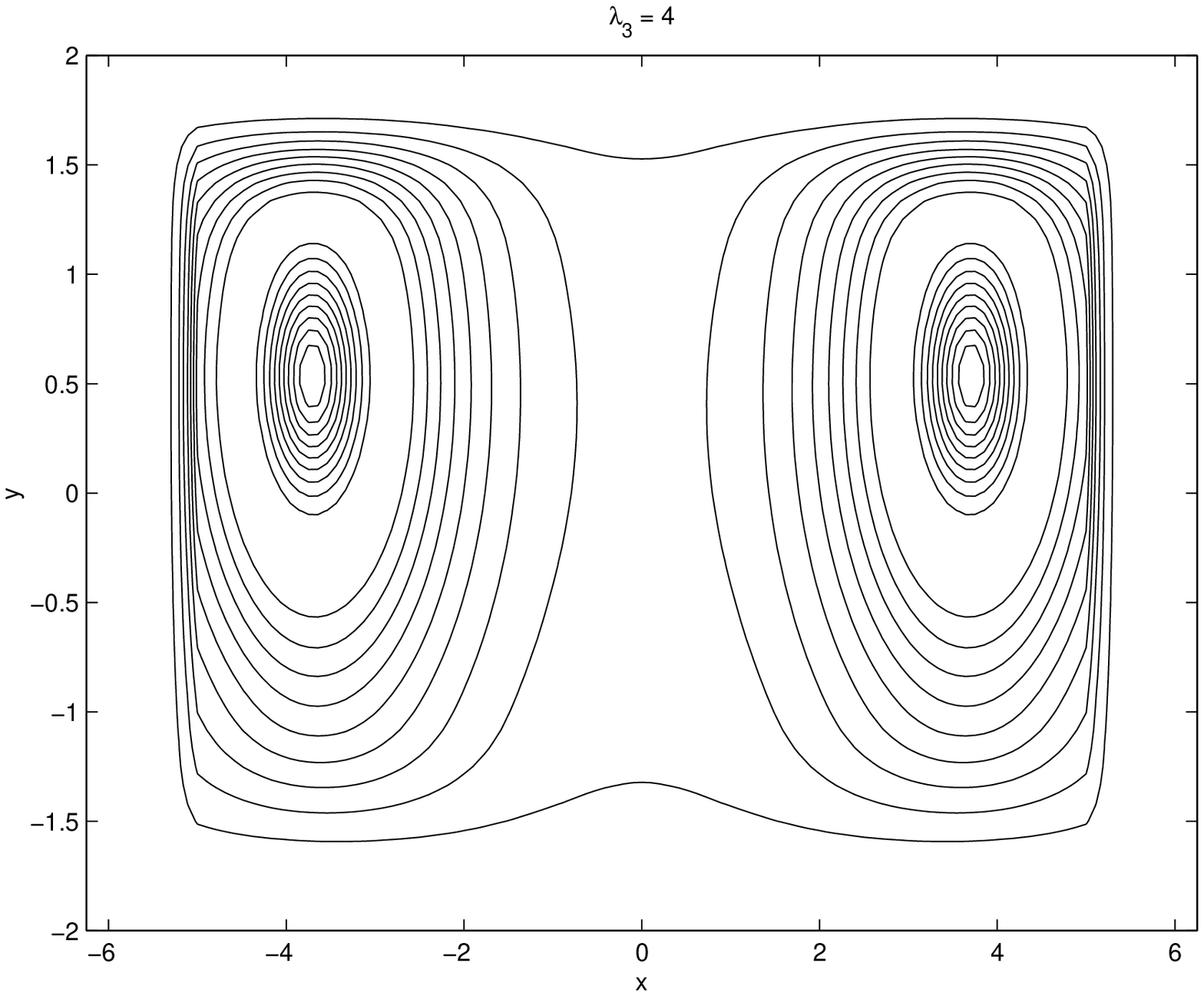}
}
\quad
\subfigure[$\l_3=1.15$: merging rings]{
\includegraphics[scale=0.4]{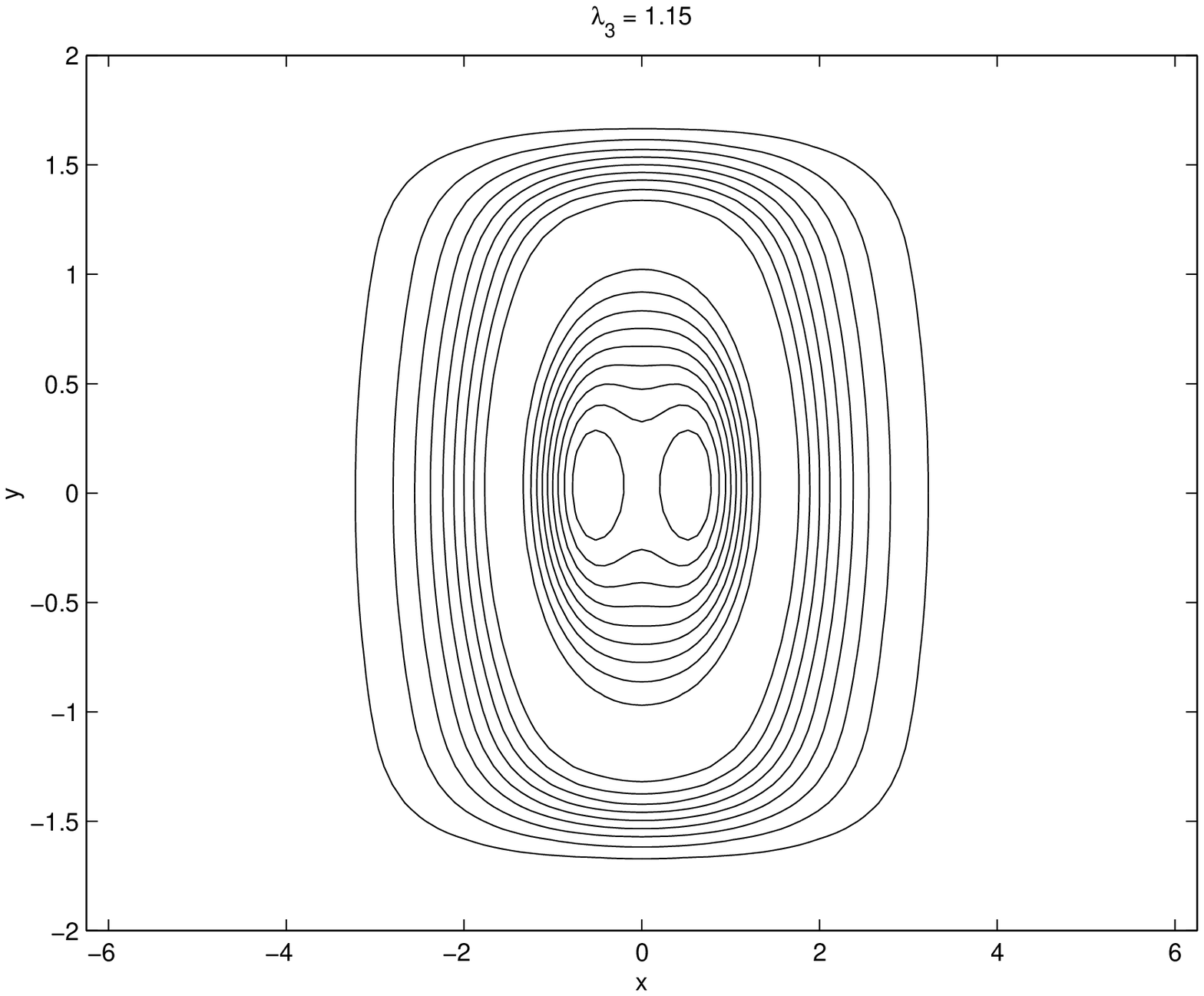}
}
\quad
\subfigure[$\l_3=1$: a single ring]{
\includegraphics[scale=0.4]{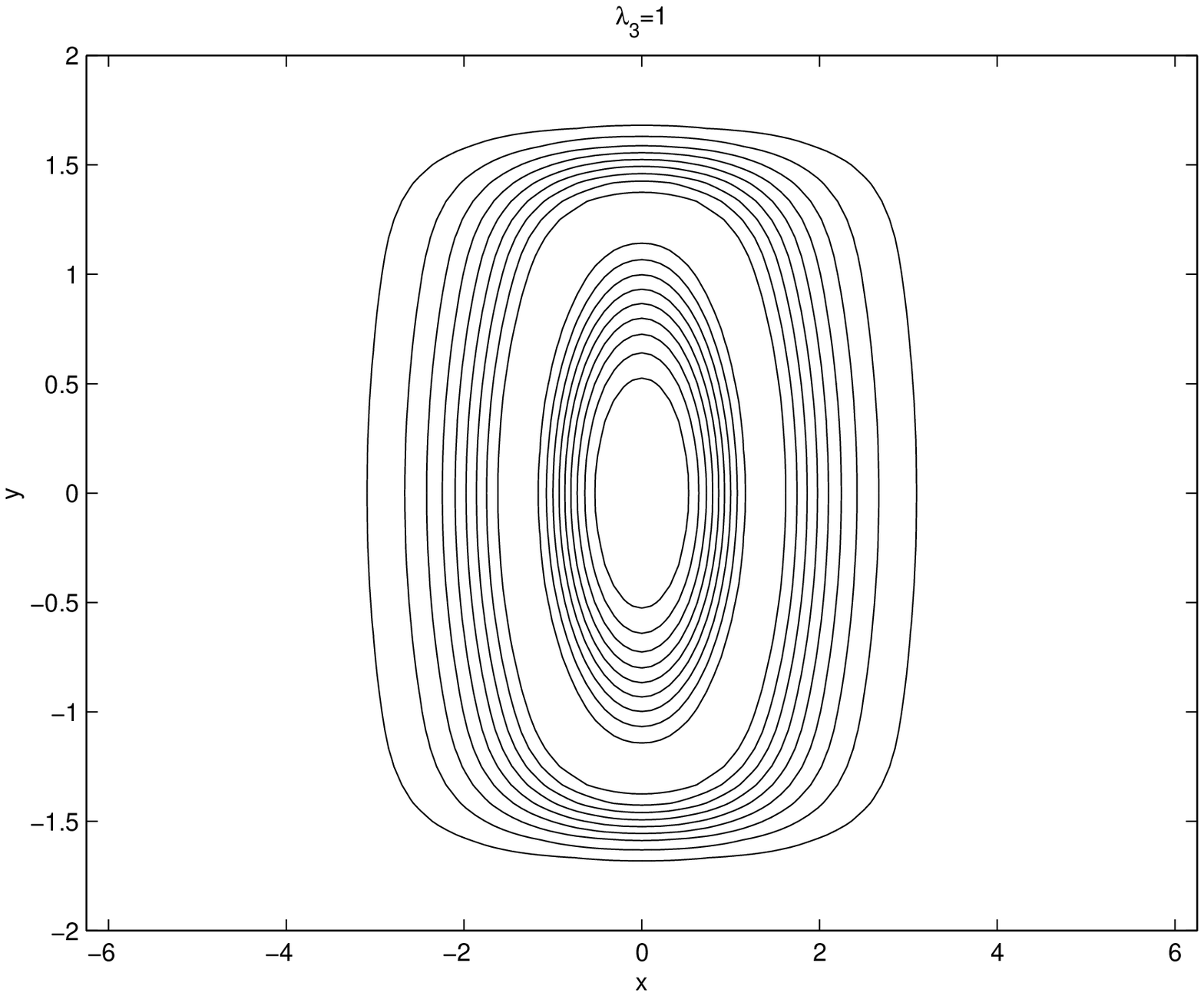}
}
\caption{A one-parameter family of soliton configurations,
  corresponding to two co-planar solitons attracting each other
  and merging to form a single $m=2$ axisymmetric configuration.
  The energy is plotted in (a), as a function of $\l_3$;
  the dotted line is the energy $2E_1$ of two widely-separated solitons.
  The other three subfigures are contour plots of $\phi^3$,
  in the $xz$-plane (the rings in (b) and (d) are circular, but look distorted
  because of the different scales on the two axes).\label{fig2}}
\end{center}
\end{figure}

Next,
consider two solitons which are far apart and co-planar (rather than co-axial
as in the previous section).  For appropriate orientations, they will
attract each other.  As the configuration moves down the energy gradient,
the solitons approach each other, and the soliton rings eventually merge
to form the single ring described in the previous paragraph.  There is a
family of instanton-generated configurations which illustrates this; the
corresponding Skyrmion picture was given in \cite{HGOA90}.  We set
$\l_1=\l_2=1$; the family is parametrized by $\l_3\in[1,\infty)$, with
$\l_3\gg1$ corresponding to the two solitons being far apart.  The poles
$X_a$ all lie in the $xy$-plane, so $t_a = z_a = 0$.  We take
\begin{equation} \label{chanb}
  x_3 = 0, \quad y_3 = -S, \quad x_2 = -x_1 = S\cos\th
     \quad y_2 = y_1 = S\sin\th,
\end{equation}
where $\th$ is determined by $\sin\th = 1/(1+\l_3)$; and where, for each
value of $\l_3$, we find the value of $S$ which minimizes the energy $E$.
Note that for $\l_3=1$, (\ref{chanb}) reduces to (\ref{equilat}).
For $\l_3\gg1$, the configuration consists of two rings in the $xy$-plane
centred on the $x$-axis at $x\approx\pm S$; this is referred to as channel B
in \cite{W00}.
The results of a numerical study are summarized in Figure 2: $E$ versus
$\l_3$ is given in Figure 2a, while the other subfigures provide contour
plots of $\phi^3$ in the $xy$-plane.  We see that the two individual
rings join to become a single ring.

Finally, we look at a one-parameter family of configurations which
interpolate between the two local minima.  For this we simply rotate
one into the other, by taking
\[\l_1=\l_2=1, \quad\l_3=\cos\psi + 0.67\sin\psi,\]
\[t_1=t_2=-0.2\sin\psi, \quad t_3=0.4\sin\psi,\]
\[x_2 = -x_1 = 1.49\cos\psi,\quad x_3=0,\]
\[y_1 = y_2 = 0.86\cos\psi,\quad y_3=-1.72\cos\psi,\]
\[z_2=-z_1=2.55\sin\psi,\quad z_3=0.\]
So when $\psi=0$, we have the (global) minimum (\ref{equilat}) with $E=2.08$;
while for $\psi=\pi/2$, we have the (local) minimum described in the
previous section, with $E=2.41$.  The energy of the configurations in this
family, as a function of $\psi$, is plotted in Figure 3.
\begin{figure}[tb]
\begin{center}
\includegraphics[scale=0.5]{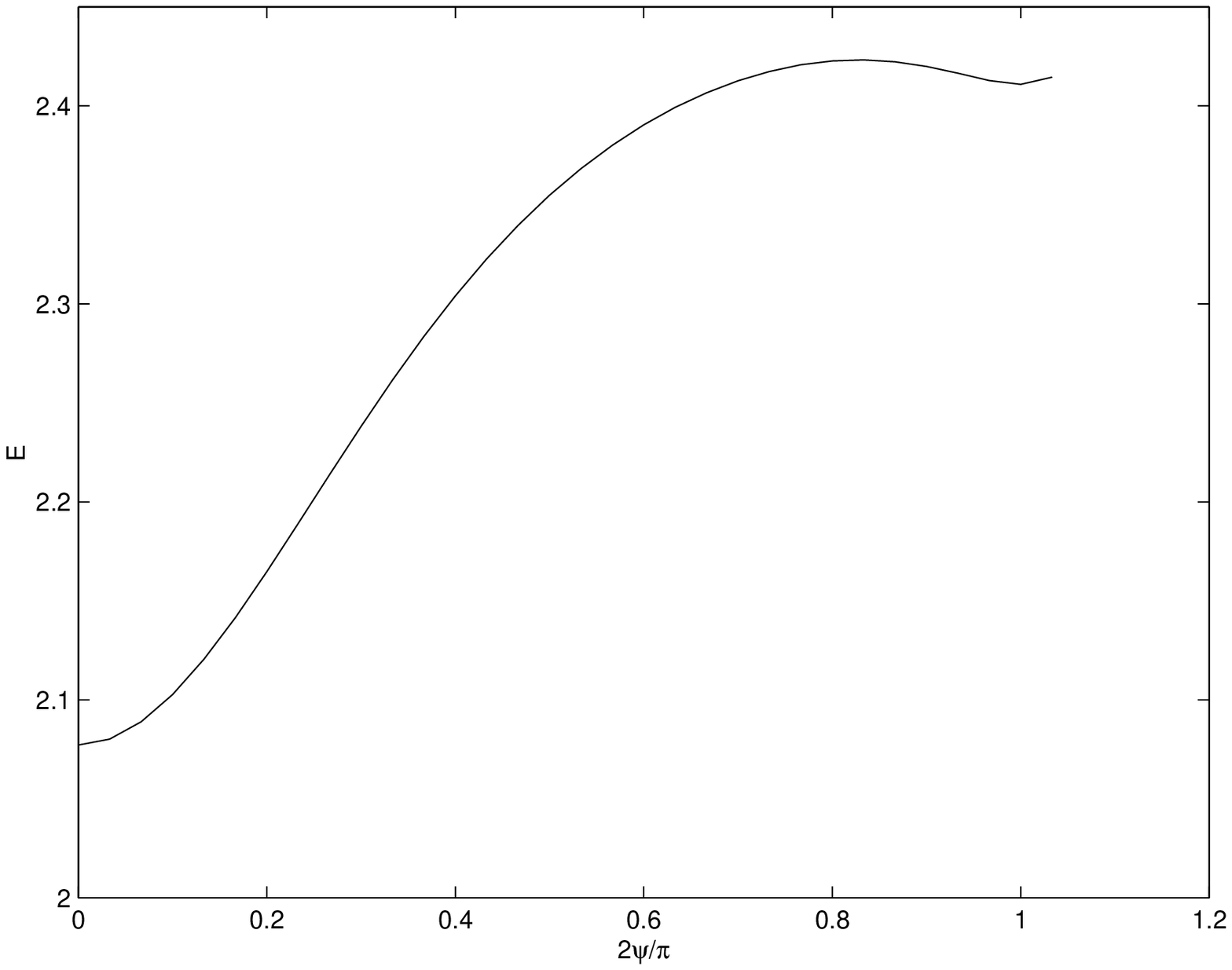}
\end{center}
\caption{The energy of a family of configurations which interpolates
 between the global $N=2$ minimum ($\psi=0$) and the other
 $N=2$ minimum ($\psi=\pi/2$). }
\label{fig3}
\end{figure}
We see that there is a path between the two minima, on which the maximum
energy is $E=2.43$ (which is less than twice the energy of two single
solitons).
In going from one minimum to the other, the curve $\phi^{-1}(0,0,-1)$
has to change from being a single ring around a double copy of the $z$-axis
(when $\psi=0$) being to a pair of rings around a single copy of the $z$-axis
(when $\psi=\pi/2$).  So the topological behaviour of the field, as one moves
along the path, is rather complicated.


\section{Concluding remarks}

In the case of the Skyrme model, the instanton approximation has been used
to study the interaction of two Skyrmions \cite{HGOA90,AM93}; and the
vibrational modes and related semiclassical quantization of the $N$-Skyrmion
for $N=2$ and $N=3$ \cite{Wal95,Wal96,H99}.  For higher $N$, the
minimal-energy Skyrmions have the appearance of various symmetric solids
(see, for example, \cite{BS01a,BS01b}), and these are quite well
approximated in terms of the ``rational map ansatz'' \cite{HMS98},
and variants thereof.  This ansatz cannot, however, provide a full
collective-coordinate manifold --- its relevance is to the description of
static, ``superimposed'' Skyrmions.

For large-$N$ Hopf solitons, one gets complicated topological structures,
with evidence of many local minima (and of large changes in the field
which do not change the energy much); see \cite{BS99,HS99,HS00}.
It seems unlikely that the instanton picture can capture all this structure
(although it does give both the minima in the $N=2$ case).
Possibly some version of the rational map ansatz might be appropriate
(it was used in \cite{BS99} to generate initial configurations for
$N\leq8$); but this remains an open question.

In this paper, we have seen that the ``space of two Hopf solitons'' can be
fairly well approximated in terms of the holonomy of Yang-Mills instantons.
The main application of this (as in the Skyrme case) is towards
understanding the dynamics of the low-$N$ soliton systems. Hopf solitons are
not spherically-symmetric --- this leads to their interactions being,
in some ways, more complicated than that of Skyrmions or monopoles.
Much further work remains to be done towards understanding their
dynamics, and the collective-coordinate description derived from
instanton holonomy might be useful in that regard.


\end{document}